# Dual-Stage Low-Complexity Reconfigurable Speech Enhancement

Jun Yang (IEEE Senior Member) and Nico Brailovsky

Facebook Reality Labs, 1 Hacker Way, Menlo Park, CA 94025, USA

*Abstract* - **This paper proposes a dual-stage, low complexity, and reconfigurable technique to enhance the speech contaminated by various types of noise sources. Driven by input data and audio contents, the proposed dual-stage speech enhancement approach performs a coarse and fine processing in the first-stage and second-stage, respectively. In this paper, we demonstrate that the proposed speech enhancement solution significantly enhances the metrics of 3-fold QUality Evaluation of Speech in Telecommunication (3QUEST) consisting of speech mean-opinion-score (SMOS) and noise MOS (NMOS) for near-field and far-field applications. Moreover, the proposed speech enhancement approach greatly improves both the signal-to-noise ratio (SNR) and subjective listening experience. For comparisons, the traditional speech enhancement methods reduce the SMOS although they increase NMOS and SNR. In addition, the proposed speech enhancement scheme can be easily adopted in both capture path and speech render path for speech communication and conferencing systems, and voice-trigger applications.**

## I. INTRODUCTION

Speech enhancement technologies have found wide applications in home audio devices (e.g., Portal from Facebook family devices), mobile audio devices (e.g., smart phones), wearable audio devices (e.g., hearing aids, smart glasses and watches), and so on. The speech enhancement aims to objectively and subjectively improve the quality and intelligibility of speech contaminated by various types of background noises [1 - 11].

Traditional speech enhancement methods usually assume that background noise is stationary and hence cannot suppress all other types of background noises in an efficient way. Also, some existing speech enhancement algorithms need the speech presence detection and estimate the background noise during speech pause. Hence, their performance greatly depends on the efficiency and accuracy of the speech presence detection. In fact, speech presence detection is a very difficult and still ultimately unsolved problem [12, 13] for realistic situations where the speech, noise, and acoustic environment are of complex and time-varying characteristics or where the desired speech and the background noise are simultaneously present with low SNR.

Challenging conditions for speech enhancement also include but are not limited to the low SNR, fast-varying probability distributions, nonlinear combinations of different types of noises, etc. In facing these adverse conditions, the existing speech enhancement methods can result in speech distortion and artifacts, fluctuate the residual noises although they have improved SNR. More importantly, these existing algorithms fail to provide the required reduction accuracies when the desired speech and the noise simultaneously present, which is the case in many practical applications.

Therefore, cost-effective and robust speech enhancement techniques that can handle more accurately and more efficiently various types of background noises are highly desirable in order to provide the necessary performance improvement for applications related to voice-trigger, speech communication and conferencing. It is the goal of this paper to provide such a desired speech enhancement solution as described in the next sections. The proposed dual-stage speech enhancement solution is trained by several hundreds of hours 16 kHz audio data recorded in real environment. It costs only ~1% of CPU on an i7-8565U core with algorithm latency less than 16 msec for frame size 8 msec, 50% overlap, and 256-point fast Fourier transform (FFT) which are used in Section III, while the existing real-time speech enhancement methods normally introduce 30 msec to 40 msec latencies [14, 15] with costing ~4% of CPU on the said core [15]. Another advantage of the proposed speech enhancement algorithm is the effective suppression of non-stationary noise due to the employment of a novel noise tracking scheme. The dual-stage speech enhancement architecture proposed in this paper can help improve both objective and subjective speech performance. Moreover, preservation of the music performance for multimedia applications can be achieved by using the proposed dynamical suppression on the basis of audio data-and-contents.

The rest of this paper is organized as follows. Section II presents the working principles and the detailed processing steps of the proposed dual-stage speech enhancement algorithm. Section III shows the objective and subjective evaluation results by the realistic test dataset and demonstrates that the proposed speech enhancement solution significantly improves the near-field SMOS and NMOS, far-field SMOS and NMOS, SNR, and listening experience. For comparisons, the traditional speech enhancement method degrades the SMOS in order to achieve the similar SNR improvement (SNRI). Section III also describes the subjective listening experience for both low SNR and high SNR use cases and demonstrates that the trained listeners obviously prefer the proposed speech enhancement algorithm to the traditional speech enhancement method. As such, the speech communication and conferencing, and barge-in performance can be improved by the proposed speech enhancement (SE) algorithm correspondingly. As the last section of this paper, Section IV makes some conclusions and further discussions.

## II. THE PROPOSED DUAL-STAGE SE ALGORITHM

Fig. 1 illustrates the schematic diagram and architecture of the proposed dual-stage reconfigurable speech enhancement algorithm with single-channel being example but without losing generality. In other words, this system is easily extendable to the multi-channel use cases. The proposed speech enhancement solution can be applied for both capture path and voice render path.

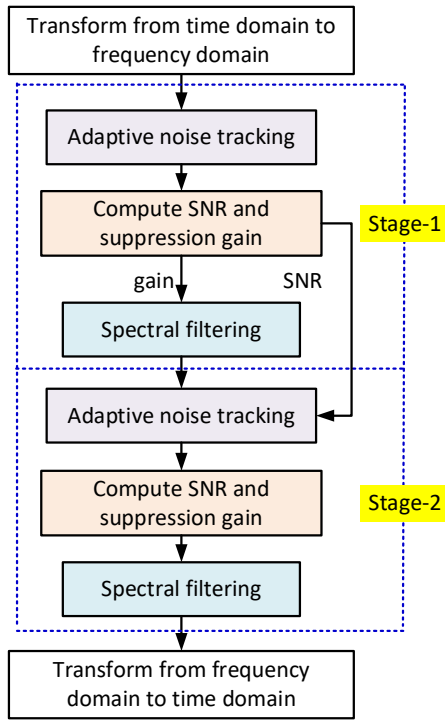

Fig. 1: The proposed dual-stage reconfigurable speech enhancement algorithm.

As shown in Fig. 1, the input signal is firstly transformed from time domain (TD) to frequency domain (FD) by some pre-processors which include a high-pass filter (HPF), input overlap, analysis windowing, FFT, and power spectral density calculation. The HPF cutoff frequency (e.g., 100 Hz in Section III), the percentage of input overlap, types of analysis window functions, and the FFT length ($N$) are reconfigurable for narrowband, wideband, and super-wideband systems according to the type of applications (such as voice-trigger, or speech communication and conferencing).

In Fig. 1, the corresponding three blocks in Stage-1 and Stage-2 are of the same algorithms and could use different parameter settings. By using the proposed dual-stage architecture, the proposed speech enhancement algorithm significantly reduces the computational complexity because it eliminates a "Transform from FD to TD" block (consisting of a synthesis window, overlap-and-add, and an IFFT processing) and a "Transform from TD to FD" block (consisting of an analysis window, overlap, and an FFT processing). Moreover, with the coarse and fine processing, the dual-stage approach achieves more SNRI and more 3QUEST improvement than the single-stage approach because Stage-2 has higher input-SNR than Stage-1 and can also make use of the SNR estimated in Stage-1 to control the Stage-2's "Adaptive noise tracking" in the way that the lower the estimated SNR, the faster is the noise tracking in Stage-2.

The "Adaptive noise tracking" blocks in Fig. 1 firstly perform bin-to-band partition to reduce computational complexity on the basis of the psychoacoustic critical bands, then search the minimum statistics in each frequency band in a time sliding fashion so that background noise spectrum is continuously updated and controlled, which is an important component to enable suppressing the nonstationary noise. The number of bands ($M$) is reconfigurable for narrowband, wideband (e.g., $M$=33 in Section III), and super-wideband applications. The "Adaptive noise tracking" block in Stage-2 generates more accurate estimation of background noise level and spectral shape than that in Stage-1.

To minimize the distortion and artifact, we propose to smooth the obtained noise magnitude spectrum over time as follows.

N(m, k) = N(m-1, k) + α(k) * (N'(m, k) - N(m-1, k))   (1)

where $N'(m, k)$ is the estimated noise magnitude spectrum at the $k$-th band ($k = 0, 1, …, M-1$) and in the $m$-th frame, $N(m, k)$ is the smoothed noise magnitude spectrum, parameter $α(k)$ is a smoothing factor between 0.0 and 1.0 which can be reconfigurable differently in Stage-1 and Stage-2. Moreover, the smaller is the SNR, the larger is the α(k) in Stage-2.

An audio data-and-content driven "Compute SNR and suppression gain" approach is employed to minimize the speech distortion and the residual noise distortion. The following are the details of the proposed subband suppression gain calculation.

$$SNR(m,k) = \frac{|X(m,k)|^2}{|N(m,k)|^2} \qquad (2)$$

$$G'(m,k) = \sqrt{1 - \frac{\mu(m,k)}{SNR(m,k)}} \qquad (3)$$

where $|X(m, k)|$ is the input sub-band magnitude spectrum at the *k-th* band and in the *m-th* frame, the value of parameter $\mu(m, k)$ is between 0.0 and 1.3 and is reconfigurable on the basis of input data and audio content. For multimedia applications, noise is underestimated and $\mu(m, k)$ is between 0.0 and 1.0. For speech communication and conferencing, noise is overestimated and $\mu(m, k)$ is between 1.0 and 1.5 (e.g., 1.49 in Section III). The larger is the value of the parameter $\mu(m, k)$, the more noise is suppressed.

The values of suppression gain are required to satisfy $\lambda(k) \leq G'(m, k) \leq 1.0$. The value of parameter $\lambda(k)$ is positive and reconfigurable according to the applications. For example, the parameter $\lambda(k)$ can be configurable to be 0.5 for voice-trigger applications and be 0.178 (used in Section III) for speech communication and conferencing application.

To generate natural sounding output signal and minimize distortion and artifact, we propose to modify the obtained suppression gain as follows.

$$G(m, k) = G(m-1, k) + \gamma(m, k) * (G'(m, k) - G(m-1, k))$$
$$\text{for } 0 \leq k \leq M-1 \qquad (4)$$

where $\gamma(m, k)$ is a time-frequency dependent smoothing factor between 0.0 and 1.0. The variable $\gamma(m, k)$ is reconfigurable such that the larger is the $G'(m, k)$, the larger value is set for the $\gamma(m, k)$. The parameter $\gamma(m, k)$ can be configured differently in Stage-1 and Stage-2.

The "Spectral filtering" block converts the obtained *M* band-suppression-gains into the *N/2* spectral-bin-gains, which are applied to the original signal spectrum by using the original phase response.

The "Transform from frequency domain to time domain" block at the end of Stage-2 in Fig. 1 performs a synthesis windowing, overlap-and-add, and an IFFT processing. A frame of signal in time domain is finally constructed with background noise suppressed to the desired degree.

To demonstrate the effectiveness of the proposed solution and the accuracy of the related analyses, two types of acoustical evaluations and related settings will be reported in the next section.

## III. EVALUATION RESULTS

We use the 3QUEST wideband measurements through the communication analysis system ACQUA [16] to objectively evaluate the speech enhancement algorithms. This is mainly because the objective evaluation metrics PESQ [17], SDR and POLQA [18] don't have a high correlation to subjective speech quality in background noise [19].

This section presents objective evaluation results of the proposed dual-stage reconfigurable SE algorithm and the traditional SE algorithm [20] in terms of near-field (NF) SMOS and NMOS, far-field (FF) SMOS and NMOS, and SNRI by using ACQUA analyses. Moreover, this section compares the subjective blind listening test results between the proposed and the traditional SE algorithms.

### A. SNRI Performance

We have conducted 3QUEST testing on ACQUA system for the following two settings of the proposed SE solution.
- NS Setting 1, i.e., SNRI is between ~13 and ~17 dB
- NS Setting 2, i.e., SNRI is between ~11 and ~15 dB

As a representative example, Fig. 2(A) and Fig. 2(B) shows that the air-conditioner noise is reduced by ~14.78 dB after our dual-stage SE processing of "NS Setting 2", that means that SNRI is about 14.78 dB in this test case.

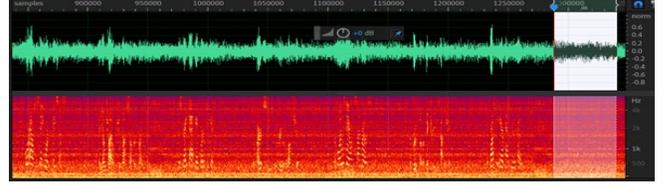

Fig. 2(A): Waveform (top) and spectrogram (bottom) before SE processing. Speech is in air-conditioner noise.

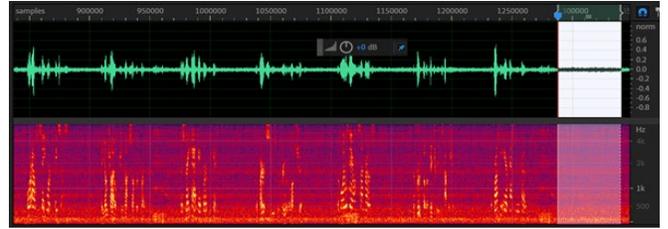

Fig. 2(B): Waveform (top) and spectrogram (bottom) after our dual-stage SE processing. Magnitude scale of waveform is the same as that in Fig. 2(A).

### B. Performance of NF and FF SMOS and NMOS

We have run the simulations of both the proposed speech enhancement algorithm and the traditional speech enhancement method by using 108 realistic test vectors consisting of 54 FF test vectors and 54 NF test vectors. Each test vector is of ~83 seconds and includes male and female speech. The dataset has covered the following test conditions:
- two distance-settings between the device-under-test (DUT) and the head-and-torso-simulator: 1 meter (i.e., near field) and 4 meters (i.e., far field),
- three input-SNRs at DUT: 0 dB, 6 dB and 12 dB,
- two speech-levels at mouth reference point: 89 dBC and 95 dBC,
- 9 types of background noises: 4 stationary noises (i.e., air-conditioner noise, fan noise, rain noise, pink noise), 4 non-stationary noises (i.e., café noise, living-room noise, office noise, Pub noise), and a rock musical noise.

Each bar in Figs. 3 and 4 denotes for the average score over 54 test vectors (54 = 3 SNRs * 2 Levels * 9 Noises).

Fig. 3 shows the relative improvements of FF SMOS, FF NMOS, NF SMOS, and NF NMOS by our proposed dual-stage speech enhancement approach. The relative improvement is calculated as follows.

$$\text{Ratio} = 100 * (MOS_a - MOS_b)/MOS_b \qquad (4)$$

where $MOS_b$ and $MOS_a$ are MOS scores before and after the speech enhancement, respectively. It can be seen from Fig. 3

that (1) our proposed dual-stage speech enhancement solution relatively improves all four types of MOS scores by 4.72%, 10.38%, 17.67%, and 30.67%, respectively, (2) FF scores are improved more than NF scores, and (3) NMOS is significantly improved more than SMOS.

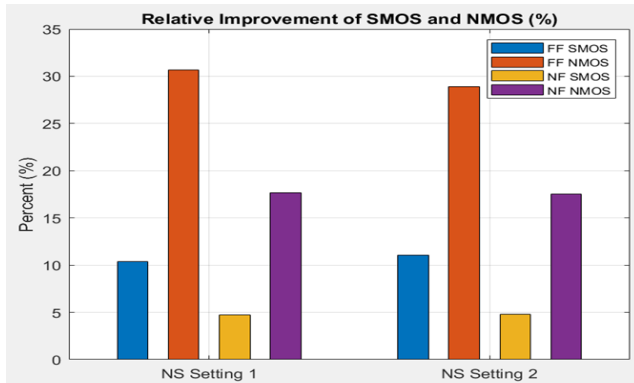

Fig. 3: Relative improvements of FF and NF SMOS and NMOS by our speech enhancement processing.

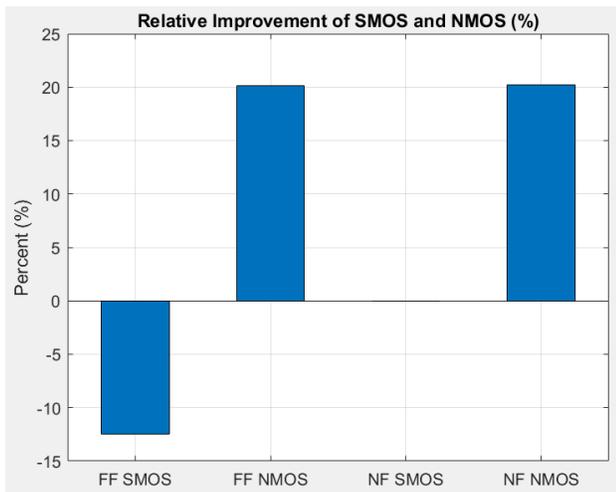

Fig. 4: Relative improvements of NF and FF SMOS and NMOS by traditional speech enhancement processing.

Fig. 4 shows the relative improvements of FF SMOS, FF NMOS, NF SMOS, and NF NMOS by the traditional speech enhancement method with SNRI ranging from 10 dB to 12 dB. It can be seen from Fig. 4 that the traditional method relatively improves the FF NMOS and NF NMOS by 20.17% and 20.21%, respectively, but relatively **degrades** FF SMOS and NF SMOS by 12.50% and 0.03%, respectively. Therefore, the proposed dual-stage approach objectively outperforms the traditional method through the apple-to-apple comparing the results in Fig. 3 and Fig. 4.

In addition to the above objective tests, the subjective listening test results have shown that the proposed dual-stage speech enhancement approach subjectively outperforms the traditional speech enhancement method in terms of speech perception in noise, speech quality, intelligibility and level, and the residual noise quality and level.

*C. Subjective Listening Testing for Two Use Scenarios*

We have run the simulations of both the proposed speech enhancement algorithm and the traditional SE method for Case #1 and Case #2 with the following test conditions.

- Case #1: There are six test vectors corresponding to six types of noise (i.e., café noise, fan noise, living-room noise, office noise, pink noise, and Pub noise). Each test vector is of 85.52 seconds, includes male and female speech, and is captured from a microphone on a smart speaker with SNR between -5 dB and 5 dB.
- Case #2: There are four test vectors. Each test vector is of 59.7 seconds and is captured from a beamforming output of a prototype device with SNR ranging from 18 dB to 27 dB.

14 trained listeners have participated in the blind listening testing. They were required to do A/B comparisons and provide their preference (aka subjective MOS) according to their perceived speech quality and residual noise quality.

Table 1: Preference of the "Proposed versus Traditional" Speech Enhancement Methods

| Speech Enhancement Algo. | | Proposed | Traditional |
|---|---|---|---|
| Percentage of the Preferred | Case #1 | 87.5% | 12.5% |
| | Case #2 | 65.0% | 35.0% |

Table 2: Preference across Noise Types in Case #1

| Noise Types | Percentage of the Preferred | |
|---|---|---|
| | *Proposed SE* | *Traditional SE* |
| Café Noise | 80% | 20% |
| Fan Noise | 89% | 11% |
| Livingroom | 100% | 0% |
| Office Nose | 100% | 0% |
| Pink Noise | 56% | 44% |
| Pub Noise | 100% | 0% |

Tables 1 and 2 have shown their listening test results. It should be noted that the larger is the percentage, the better the speech enhancement performance is.

It can be seen from Table 1 that the proposed dual-stage speech enhancement algorithm obviously outperforms the traditional speech enhancement method for both scenarios. Table 2 demonstrates that the proposed dual-stage speech enhancement algorithm obviously outperforms the traditional speech enhancement method for all the six types of noises.

In summary, both objective evaluations and subjective listening testing results have demonstrated that the speech is significantly enhanced in noisy conditions by the proposed SE approach. More importantly, the processed speech and the residual noise sound better in both high quality and natural way by using our proposed speech enhancement solution.

## IV. CONCLUSIONS

Focusing on the working principles, algorithm details and testing comparisons, a new dual-stage speech enhancement solution driven by input-data and audio-content has been reported in this paper. It has been shown that the proposed scheme obviously improves all related performance indicators consisting of far field SMOS and NMOS, near field SMOS and NMOS, SNR, and subjective listening experience.

The proposed speech enhancement algorithm is easily extendable to multi-channel use cases. Example-1: it can be used as a non-localized speech enhancement (aka post-filtering) in the signal beam generated by a localized speech enhancement (e.g., beamforming) in a multi-microphone speech enhancement system. In addition, the noise estimation and suppression can also be performed by using the noise beam to achieve the further noise suppression. Example-2: the proposed algorithm can be used in a stereo-voice render path where each channel uses the same or independent settings and is applied by the corresponding left and right suppression gain or applied by the maximum suppression gain in each frame to preserve the stereo image.

Furthermore, this proposed SE processing supports for any sampling rate and any frame size. Without introducing artifacts (e.g., musical noise) the proposed solution can be reconfigurable in such a way as to support various applications and products with different noise types and parameters settings. The related parameters can be predefined and tuned by the related corresponding training datasets.

ACKNOWLEDGMENT

We wish to express our sincere thanks to Lidia Castanon Campoy for conducting the 3QUEST testing of Fig. 4. We also express our special thanks to Joel Clark, Joshua Bingham, Christopher Evans, Marshall Chiu, Rick Chao, Mengchuan Wang, Brian Adair, Nico Brailovsky, Ethan Schreiber, Mike Petterson, J. White, Terry Cho, Richard Webb, and Johan Blome for participating in the blind listening testing.